\documentclass[%
 reprint,
 amsmath,amssymb,
aps,onecolumn
]{revtex4}
\usepackage{float}
\usepackage{graphicx}
\begin{document}
\title{Supersolid phases of dipolar fermions in a 2D lattices bilayer array}

\author{A. Camacho-Guardian and R. Paredes}                   

\affiliation{ 
 Instituto de F\'{\i}sica, Universidad
Nacional Aut\'onoma de M\'exico, Apartado Postal 20-364, M\'exico D.
F. 01000, Mexico. }


\date{\today}
\begin{abstract}
Supersolid phases as a result of a novel coexistence of superfluid and density ordered checkerboard phases are predicted to appear in ultracold Fermi molecules confined in a bilayer array of 2D square optical lattices. We demonstrate the existence of these phases within the inhomogeneous mean field approach. In particular, we show that tuning the interlayer separation distance at a fixed value of the chemical potential produces different fractions of superfluid, density ordered and supersolid phases.

\end{abstract}

\maketitle
\section{Introduction}
\label{intro}
Several phases of matter appear in the quantum degenerate regime only, namely, superconductivity, superfluidity (SF) and supersolid (SS) phases. While the two formers have been successfully explained, the supersolid phase \cite{Kim, Choi}, consistent with thermodynamic stability criteria \cite{Mendoza}, remains as a subject of theoretical investigation. On the other hand, not at necessarily low temperatures, but also manifesting many body quantum statistical behavior, the High-$T_c$ (HTc) superconductivity phenomenon still continues as an open question in the context of condensed matter.
Experiments with ultracold neutral gases are at the present time the closest candidates to quantum simulate, and thus address the description, of such not quite yet understood quantum phases \cite{Hofstetter, Chan, Liu, Buhler, Wang, Fujihara}.  

Recent experimental studies have shown how quantum phases as SF, SS, Mott insulator and charge density wave, emerge from competing short- and long-range interactions among ultracold Bose atoms confined in an optical lattice coupled to a high finesse optical cavity \cite{Esslinger}. Those phases arise as a result of exploiting the matter-light coupling since in such a case interactions can be tuned on demand. There is however an alternative way of handling either the range and direction of interactions in ultracold neutral gases, which is by confining dipolar atoms or molecules in optical lattices. As it has been shown from the theoretical perspective, the combination of both, the long range anisotropic character of dipolar interactions and the controllable lattice structure where the atoms/molecules lie, make the many-body physics becomes very rich \cite{Baranov2, Lahaye, Chen, Zinner}.

\begin{figure}[htbp]
\begin{center}
\includegraphics[width=3.0in]{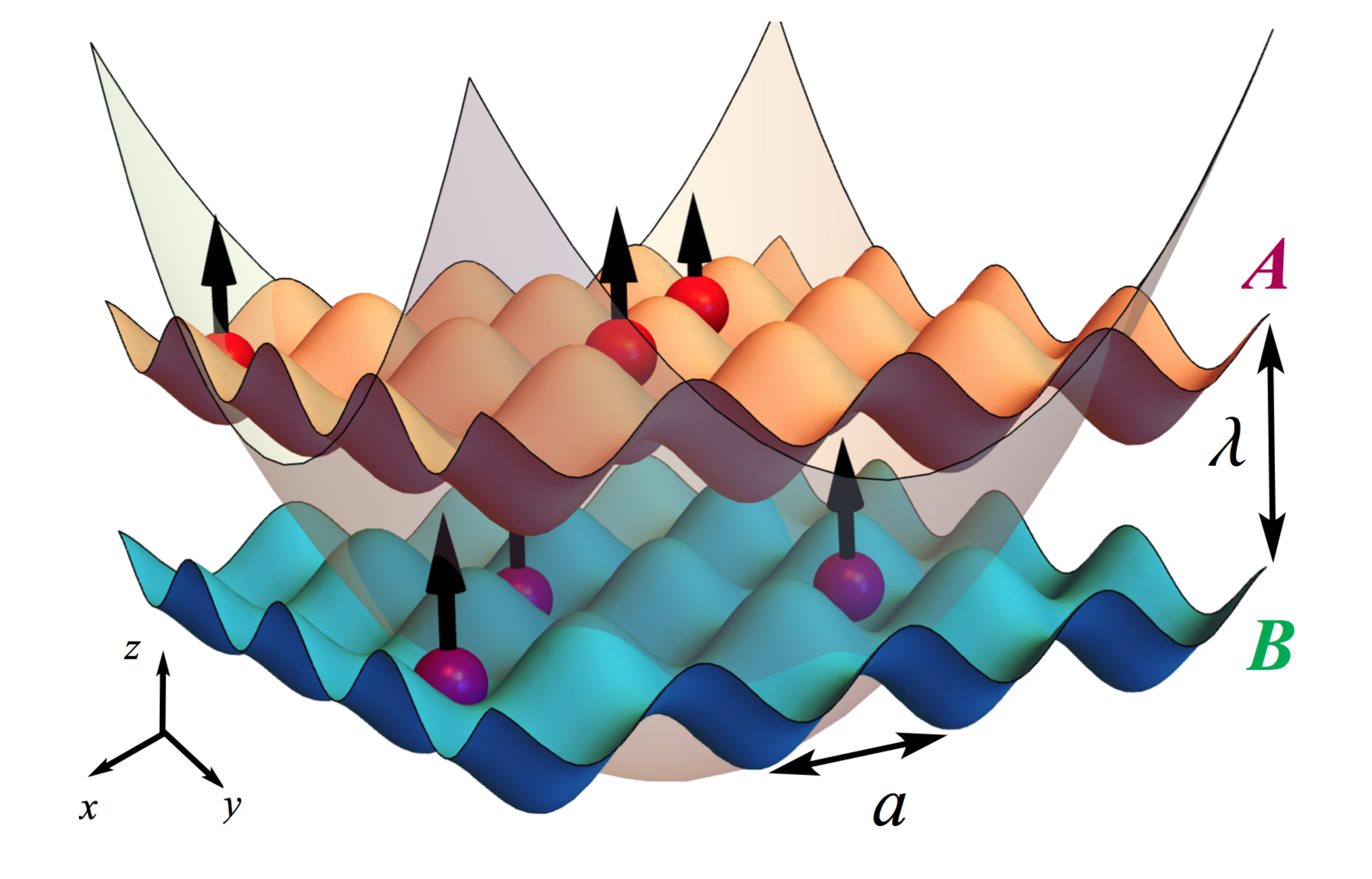} 
\end{center}
\caption{(Color online) Schematic representation of the dipolar Fermi gas. The molecules affected by a harmonic trap potential lie in the lattice sites in up and down layers.}
\label{Fig1}
\end{figure}

In this work we consider a model proposed previously \cite{Vanhala, Camacho, Ancilotto} to demonstrate that ordered density wave (DW), SS and SF phases can be accessed by changing the external fields that set the system. In Fig. \ref{Fig1} we show a scheme of the quantum simulator that can be created in the laboratory to explore the referred phases. The model system is composed of dipolar Fermi molecules lying in a bilayer array of square lattices in 2D. Although such a configuration has not been realized yet, the current experimental panorama of ultracold dipolar gases, in particular the potential capacity of loading long-lived Fermi molecules of NaK, KRb and NaLi \cite{Woo, Ni} in optical lattices as well as the recently produced ro-vibrational ground state in molecules of NaK, is promising in setting the array here considered.

Previous mean field analysis on dipolar fermions placed onto a single-layer square lattice, with arbitrary orientation and considering dipoles with fixed orientation, have predicted the melting among SF and DW phases \cite{Zinner1, Gadsbolle1, Gadsbolle2, Bhongale} and a variety of DW phases \cite{Mikelsons} respectively. Also, an extended model including a mixture of Fermi molecules with contact interactions, loaded in a bilayer array predicted density ordered phases as well as superfluids phases \cite{Vanhala, Prasad, Baranov, Pikovski, Potter}. The possibility of supersolid phases in these dipolar Fermi gases has also been studied \cite{He,Gadsbolle1,Gadsbolle2}. On the other side SF, Mott insulating, DW and SS phases of He have been investigated within a mean field context too \cite{Rica,Ye,Nozieres}. In the present study we consider dipolar Fermi molecules situated in a double array of parallel optical lattices having dipole orientation perpendicular to the lattice in the presence of a harmonic trap, to demonstrate that in addition to SF and DW patterns there is a region of coexistence in the phase diagram where SS phases emerge. Working within the Bogoliubov-de Gennes (BdG) approach we show that depending upon carefully controlled parameters these phases can be accessed under current experimental conditions.

The paper is organized as follows. In Section \ref{Model} we introduce the model considered in our study and describe the theoretical approach employed. In Section \ref{SS} we illustrate the coexistence and spatial overlap among superfluid and DW phases for several values of the temperature. We summarize our findings in the phase diagram at finite and zero temperature. Finally, we present our conclusions in Section \ref{Conclusion}.

\section{Model}
\label{Model}
We consider Fermi molecules of dipole moment $d$ and mass $m$ lying in two parallel square lattices of lattice constant $a$ separated by a distance $\lambda$, and a harmonic trap with frequency $\omega$ (see Fig. \ref{Fig1}). In the presence of an electric field perpendicular to the layers, the dipoles align along the same direction. Fermions in the same layer repel each other always, however, dipoles in different layers attract each other at short range, while also repelling each other at large distances. Thus, interaction between fermions in the same and in different layers is given respectively by
\begin{eqnarray}\nonumber
 &&V^{\alpha,\alpha}(\vec{r})=d^2\frac{1}{r^3},\\
&&V^{\alpha,\beta}(\vec{r})=d^2\frac{r^2-2\lambda^2}{(r^2+\lambda^2)^{5/2}},
\label{V_int}
\end{eqnarray}
where $r$ is the in-plane distance between two fermions. Greek indices label the layer where the molecule is placed. Thus, superscripts $\alpha,\alpha$ ($\alpha,\beta$) indicate that interaction occurs between fermions in same (different) layers. For clarity, we denote the intralayer interaction by $V^{\alpha,\alpha}(\vec{r})=V(\vec{r})$, and the interlayer interaction by $V^{\alpha,\beta}(\vec{r})=U(\vec{r})$.
The system is described by the Hubbard model with the Hamiltonian given by $\hat{H}=\hat{H}_0+\hat{V}+\hat{U}$, with{\small
\begin{eqnarray} \nonumber
&& \hat{H}_0=\sum_{\alpha=A,B}\sum_{\vec{k}}(\epsilon_{\vec{k}}-\mu_\alpha)\hat{n}_{\vec{k}}^{\alpha}+\sum_{\alpha=A,B}\sum_{\vec{i}}\frac{m\omega^2}{2} r^2(i)\hat{n}_{\vec{i}}^{\alpha}\\ \nonumber
&&\hat{V}=\frac{1}{2\Omega}\sum_{\alpha=A,B}\sum_{\vec{k},\vec{k'},\vec{q}}V(\vec{q})\hat{c}^\dagger_{\vec{k}+\vec{q},\alpha}\hat{c}_{\vec{k},\alpha}\hat{c}^\dagger_{\vec{k'}-\vec{q},\alpha}\hat{c}_{\vec{k'},\alpha}\\ \nonumber
&&\hat{U}=\frac{1}{\Omega}\sum_{\vec{k},\vec{k'},\vec{q}}U(\vec{k}-\vec{k'})\hat{c}^\dagger_{\vec{q}/2+\vec{k},A}\hat{c}^\dagger_{\vec{q}/2-\vec{k},B}\hat{c}_{\vec{q}/2-\vec{k'},B}\hat{c}_{\vec{q}/2+\vec{k'},A}\\
\label{H}
\end{eqnarray}}
where $\hat{c}^\dagger_{\vec{k},\alpha}, \hat{c}_{\vec{k},\alpha}$  are the standard creation and annihilation operators and $\hat{n}_{\vec{k},\alpha}=c^\dagger_{\vec{k},\alpha}c_{\vec{k},\alpha}$, $\epsilon_{\vec k}= -2t(\cos{k_x a}+ \cos{k_y a})$ is the in-plane energy dispersion of the ideal Fermi gas within the tight binding approximation, being $t$ the hopping among nearest neighbors. $V(\vec q)$ and $U(\vec k - \vec k')$ are the Fourier transforms of $V(\vec r- \vec r')$ and $U(\vec r)$, respectively and $\Omega$ the number of sites. The terms containing the harmonic confinement are written in the Fock basis of sites, where the vector position in the lattice is denoted by $\vec{r}(i)=a(i_x,i_y)$, and $\omega$ the frequency of the harmonic trap that confines the molecules. In what follows, all the energies will be scaled with respect to $t$. We also introduce two relevant physical quantities: the dipolar interaction strength $a_d=m_{eff}d^2/\hbar^2$, with $m_{eff}= \hbar^2/2ta^2$ the effective mass, and the dimensionless parameters $\Lambda= \lambda/a$ and $\chi=a_d/a$. 

The proposed model can be mapped into a system of fermions in two different hyperfine spin states $\uparrow, \downarrow$ ($A\rightarrow \uparrow, B\rightarrow \downarrow$) confined in a 2D lattice. The terms $V$ and $U$ describe repulsive and attractive interactions among fermions in the same and different hyperfine states respectively. We should stress that $U$ is an interaction that is attractive at short distances while becoming repulsive at long distances $U$. Thus, by controlling the separation between the layers $\lambda$, the proposed model represents a promising candidate to study the quantum phases from competing short- and long-range interactions as well as attractive versus repulsive interactions.  The inclusion of the harmonic potential plays also a crucial role since, as it is well known, the global thermodynamics of the phase transition is qualitatively different from that of the homogeneous case \cite{Ayala}. A remarkable signature of this fact is that the magnitude of the coherence length can be as large as the typical confinement distance.  

To investigate the physics of the model described above, we use mean-field theory including the usual BCS pairing terms and  the Hartree contributions\cite{Hartree}. We expect this approximation to be reasonably accurate in the weakly interacting regime. The mean-field Hamiltonian is diagonalized by solving the Bogoliubov-de Gennes equations (BdG) \cite{Convergence}. This mean-field approach is commonly used for studying competing magnetic and superconducting phases in the context of high $T_c$ superconductivity \cite{Chen, Chen2} and in the context of ultra cold fermions \cite{Andersen}. It also has  been recently employed  for describing  strongly correlated systems, like  effective $p$-wave interaction and topological superfluids in quantum gases in lower dimensions systems (1D and 2D)\cite{Wang}. The equation to be solved is,
$$
\sum_j\left( \begin{array}{cc}
H^{0}_{ij,\alpha} & \Delta_{i,j}  \\
\Delta_{i,j} & -H_{ij,\bar{\alpha}}^0   \end{array} \right)\left( \begin{array}{c}
u_{j,\alpha}^n\\
v_{j,\bar{\alpha}}^n \end{array} \right)=E_n\left( \begin{array}{c}
u_{j,\alpha}^n\\
v_{j,\bar{\alpha}}^n \end{array} \right),$$
where the matrix elements $H^0_{ij,\alpha}$ incorporate the tunneling among nearest neighbors $t\delta_{\langle i,j\rangle}$, the effect of the harmonic confinement $\epsilon_i=\frac{m\omega^2}{2} r^2(i)$ and the inter site interaction on the Hartree level that is expected to dominate \cite{Hartree}, that is, $H^0_{ij,\alpha}=-t\delta_{\langle i,j\rangle}+\left(\sum_{l\neq i}V_{li}\langle n_{l,\alpha}\rangle+\epsilon_i-\mu\right)\delta_{i,j}$ with $\delta_{\langle i,j\rangle}$ the Kronecker delta for nearest neighbors.  $\Delta_{i,j}=U_{i,j}\langle c_{i,A} c_{j,B}\rangle$ is the superfluid parameter. The eigenvalues denoted by $E_n$ are self-consistency obtained through the usual relations  $n_{i,A}=\sum_{n}|u_{i,A}|^2f(E_n)$ and $n_{i,B}=\sum_{n}|v_{i,B}|^2(1-f(E_n))$ with $(u^n_{i\alpha},v^n_{i,\bar{\alpha}})$ the local Bogoliubov quasiparticle amplitudes, being $f(E_n)$ the Fermi distribution and $n_{i,\alpha}$ the expectation value of $\hat{n}_{i,\alpha}$. For simplicity we shall assume that both layers are equally populated.

To include the effects of the harmonic trap, in addition to the usual order parameters that globally describe the system, we introduce local order parameters. The global density order parameter is given by $\rho_{\vec{Q},\alpha}=\frac{1}{N_\alpha}\sum_{\vec{k}}c^\dagger_{\vec{k}+\vec{Q}}c_{\vec{k}}$, where the vector $\vec{Q}$ identifies either, a checkerboard pattern $\vec{Q}=(\pi/a,\pi/a)$, or a stripe density order $\vec{Q}=(\pi/a,0),$ or $ (0,\pi/a)$. The local density order parameter is given by $\phi_{i}=\sum_{j_i} (-1)^{j_x+j_y}n_j$, where the index $j_i$ denotes that the sum runs over the first and second nearest neighbors. This local order parameter describes a checkerboard density pattern in a $3\times 3$ sublattice centered at site $i$. The superfluid local order parameter is given by $\Delta_i=\sum_{j_{i}}\Delta_{i,j}$, being the average superfluid behavior studied through  $\Delta=\sum_{i}\Delta_{i}/N_A$.

\section{Supersolid: coexistence of Superfluid and Density Ordered phases}
\label{SS}
To determine the density profile and the behavior of the gap across the lattice, we solve  BdG equations for lattices of size $\Omega = 2 \times 37 \times 37$, maintaining fixed the value of chemical potential $\mu/t=1.5$ \cite{Romero}. This restriction causes that the total number of fermions is increased with temperature, that is, at zero temperature the number of fermions is $N_A+N_B= 320$ while for $k_B T/t=0.5$ this value is increased to $335$. We also keep fixed the values of the interaction strength and the harmonic frequency at $\chi=0.3$ and $\frac{1}{2}m(\omega a)^2/t =0.025$, respectively.

To illustrate the competition among different phases we have selected the cases $\Lambda=0.8, 0.85, 0.9 $ and $1.0$. We plot the density profile and the gap parameter profile for several values of the temperature. In particular, we chose $k_B T/t=0.0, 0.10, 0.25$ and $0.45$.

\begin{figure}[htbp]
\begin{center}
\includegraphics[width=5.0in]{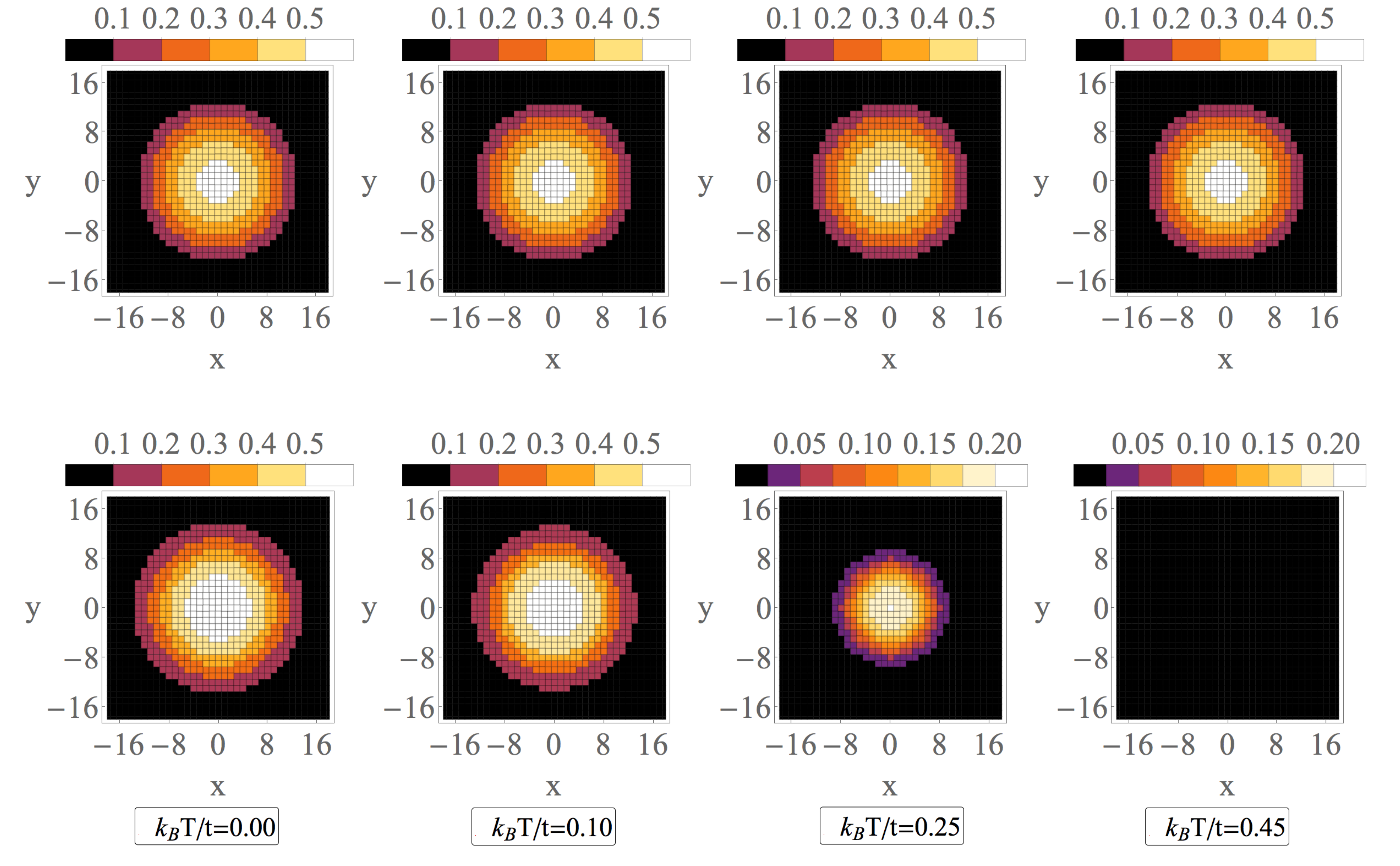} 
\end{center}
\caption{(Color online) Density order profile (top) and superfluid order parameter (bottom) as a function of the temperature. From left to right $k_B T/t=0.0, 0.10, 0.25$ and $0.45$. For $\Lambda=0.80$ and $\chi=0.3$}
\label{Fig2}
\end{figure}

First we start considering $\Lambda=0.8$. From Fig.\ref{Fig2} we observe an homogeneous distribution at the center of the trap at zero temperature for both, density and gap profiles. However, at finite temperature, the distribution of the density profile remains homogeneous at the center of the trap, while the gap structure shows a decreasing ratio as temperature is increased until they vanishes at a temperature of $k_{B}T/t=0.28$. In previous studies \cite{Camacho, Ancilotto} reported a BCS superfluid phase in the weakly interacting regime while occurring formation of dimers in the strong interaction regime, leading those dimers to a Bose superfluid. In the present work we focus on the BCS superfluid to DW-Supersolid phase transition. Therefore, the values of the parameters are restricted to those in which the weakly interacting regime is ensured.

\begin{figure}[htbp]
\begin{center}
\includegraphics[width=5.2in]{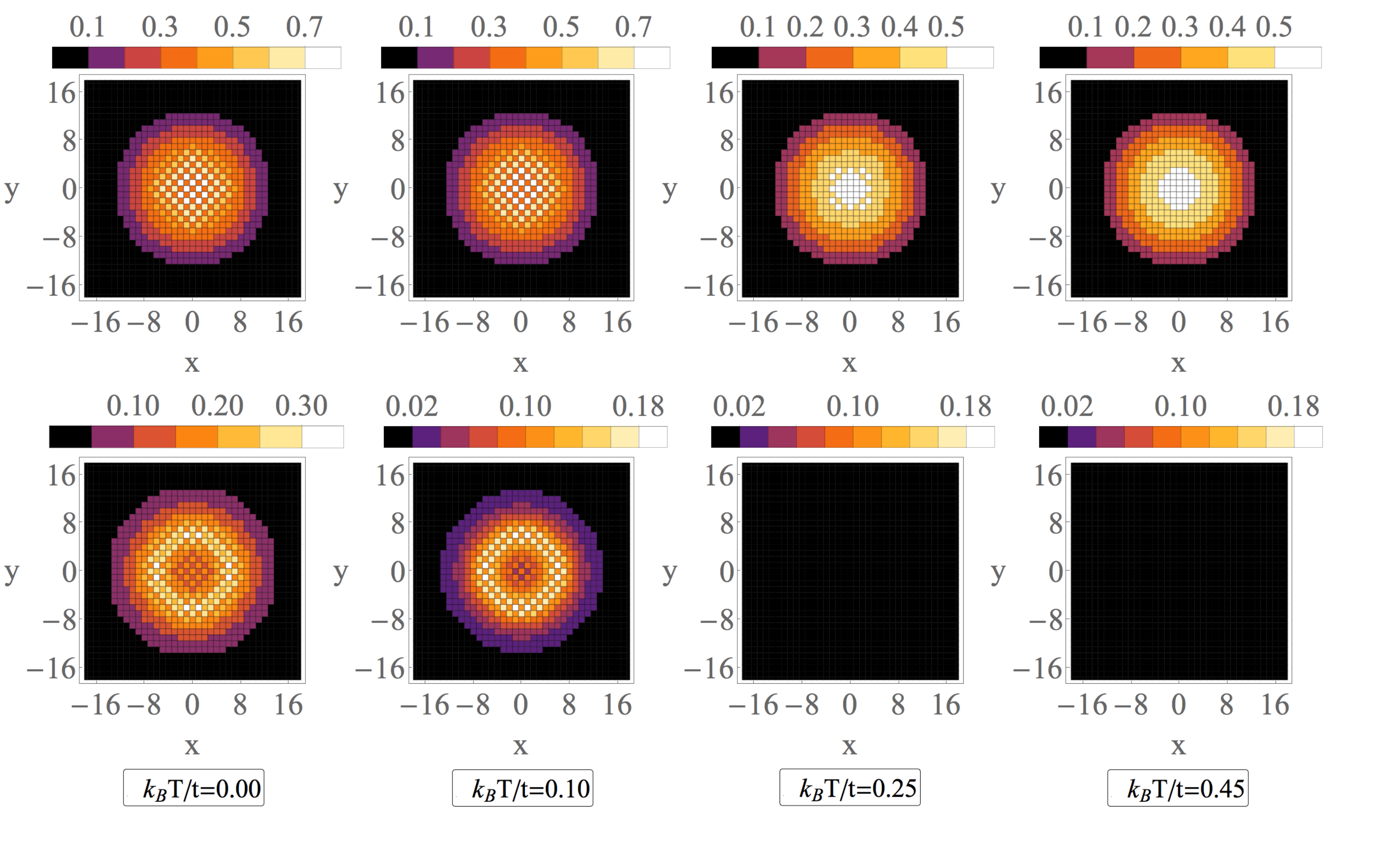} 
\end{center}
\caption{(Color online) Density order profile (top) and superfluid order parameter (bottom) as a function of the temperature. From left to right $k_B T/t=0.0, 0.10, 0.25$ and $0.45$.  For $\Lambda=0.85$ and $\chi=0.3$.}
\label{Fig3}
\end{figure}

\begin{figure}[htbp]
\begin{center}
\includegraphics[width=5.0in]{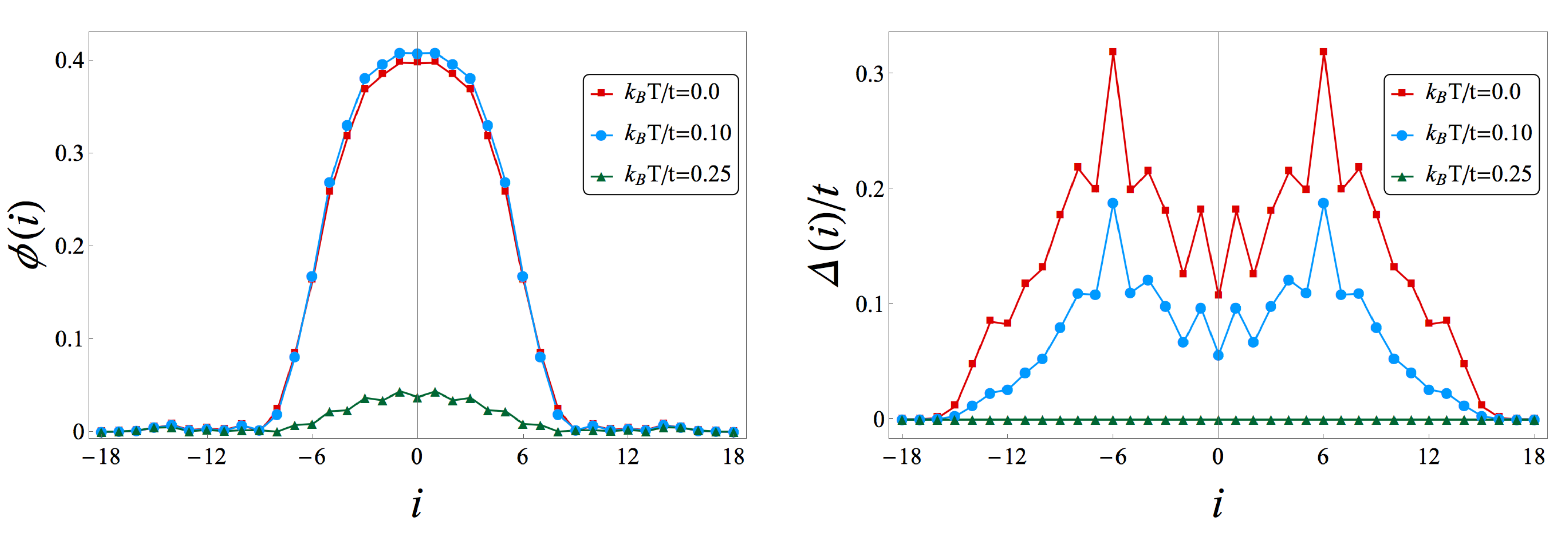} 
\end{center}
\caption{(Color online) Local order parameters. We plot a cross section of the local density $\phi(x,0)$ and the local gap $\Delta(x,0)$ through the center of the trap for $k_B T/t=0.0, 0.10, 0.25$ and $0.45$. For $\Lambda=0.85$ and $\chi=0.3$.}
\label{Fig4}
\end{figure}

When the interlayer spacing is increased, the competition between attractive and repulsive dipole interactions becomes evident. As plotted in Fig. \ref{Fig3}, for $\Lambda=0.85$ at low temperatures, there is a large region in the center of the trap having a superfluid order parameter coexisting with a checkerboard density order. This is the signature of a supersolid phase. When the temperature is increased, the radius of the superfluid order parameter shrinks and completely vanishes at a critical temperature of $k_B T/t=0.13$, while the checkerboard phase melts at a temperature of $k_B T/t=0.26$. That is, the supersolid phase exists, in this system, for certain temperatures, as further shown below in the corresponding phase diagram. Cross sections of the local order parameters are shown in Fig.\ref{Fig4}, where it can be appreciated the presence of a supersolid phase at the center of the trap, as the spatial overlap of both superfluid and DW. 

\begin{figure}[H]
\begin{center}
\includegraphics[width=5.2in]{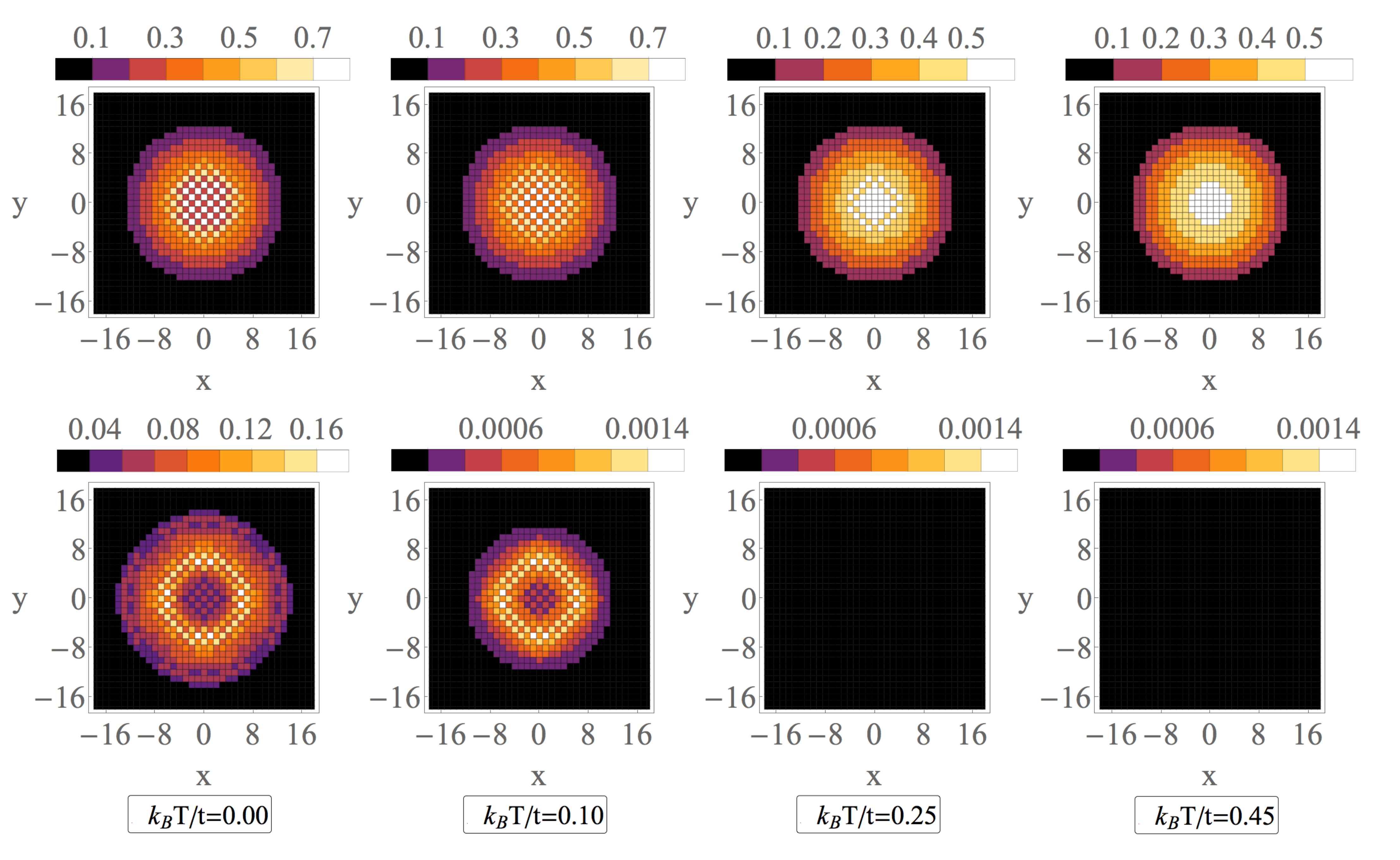} 
\end{center}
\caption{(Color online)  Density order profile (top) and superfluid order parameter (bottom) as a function of the temperature. From left to right $k_B T/t=0.0, 0.10, 0.25$ and $0.45$.  For $\Lambda=0.90$ and $\chi=0.3$.}
\label{Fig5}
\end{figure}

\begin{figure}[htbp]
\begin{center}
\includegraphics[width=5.0in]{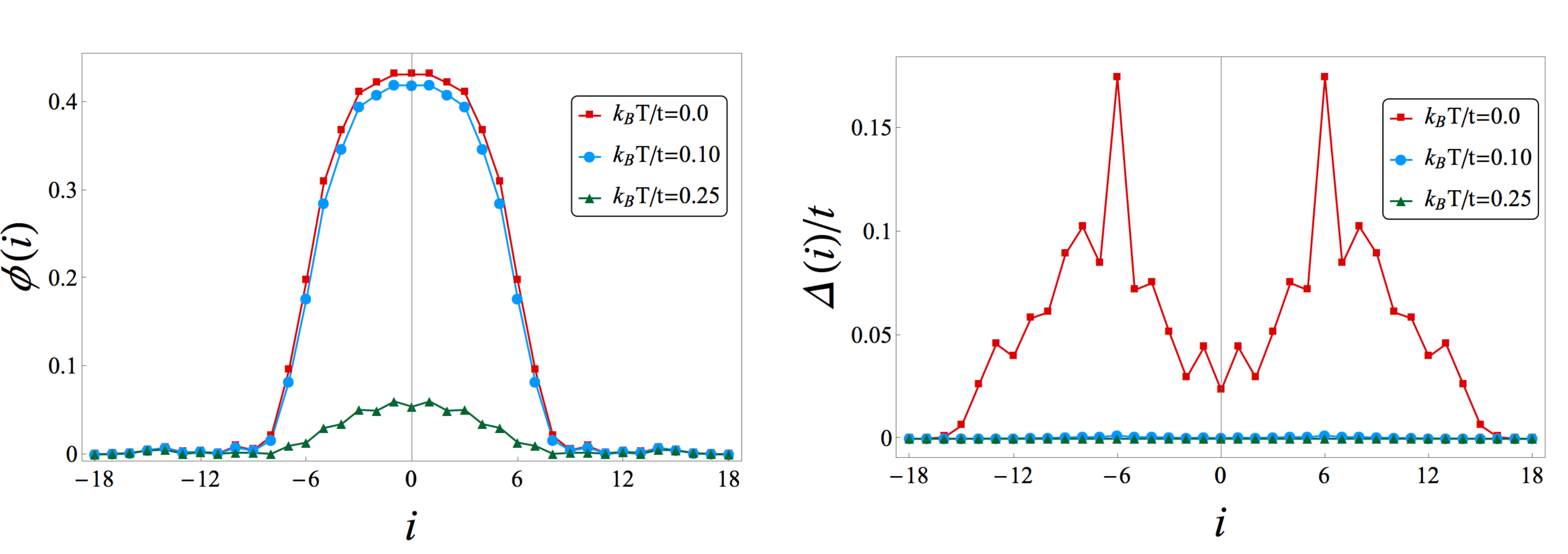} 
\end{center}
\caption{(Color online)  Local order parameters. We plot a cross section of the local density $\phi(x,0)$ and the local gap $\Delta(x,0)$ through the center for $k_B T/t=0.0, 0.10, 0.24$ and $0.45$. For $\Lambda=0.90$ and $\chi=0.3$.}
\label{Fig6}
\end{figure}

For $\Lambda=0.9$ the repulsive intralayer interaction starts to dominate at the center of the trap. From Fig. \ref{Fig5} one can see that there is a wide region at the center of the trap exhibiting a checkerboard DW pattern. Such patterns persist below temperatures of $k_B T/t=0.26$. We also observe that the superfluid order parameter appears to surround the checkerboard pattern and that such a superfluid disk completely vanishes when the temperature reaches a value of $k_B T/t=0.06$. The cross sections shown in Fig. \ref{Fig6} exhibit a small region where both phases spatially overlap. Other studies \cite{Kurdestany,Pai} with different systems in the presence of a harmonic trap have shown coexistence of phases without spatial overlapping, for instance, in the extended Bose-Hubbard model the Mott insulator and superfluid phases tend to form rings and disks. In 2D those studies agree quantitatively with Quantum Monte Carlo calculations and more sophisticated methods.

\begin{figure}[htbp]
\begin{center}
\includegraphics[width=5.2in]{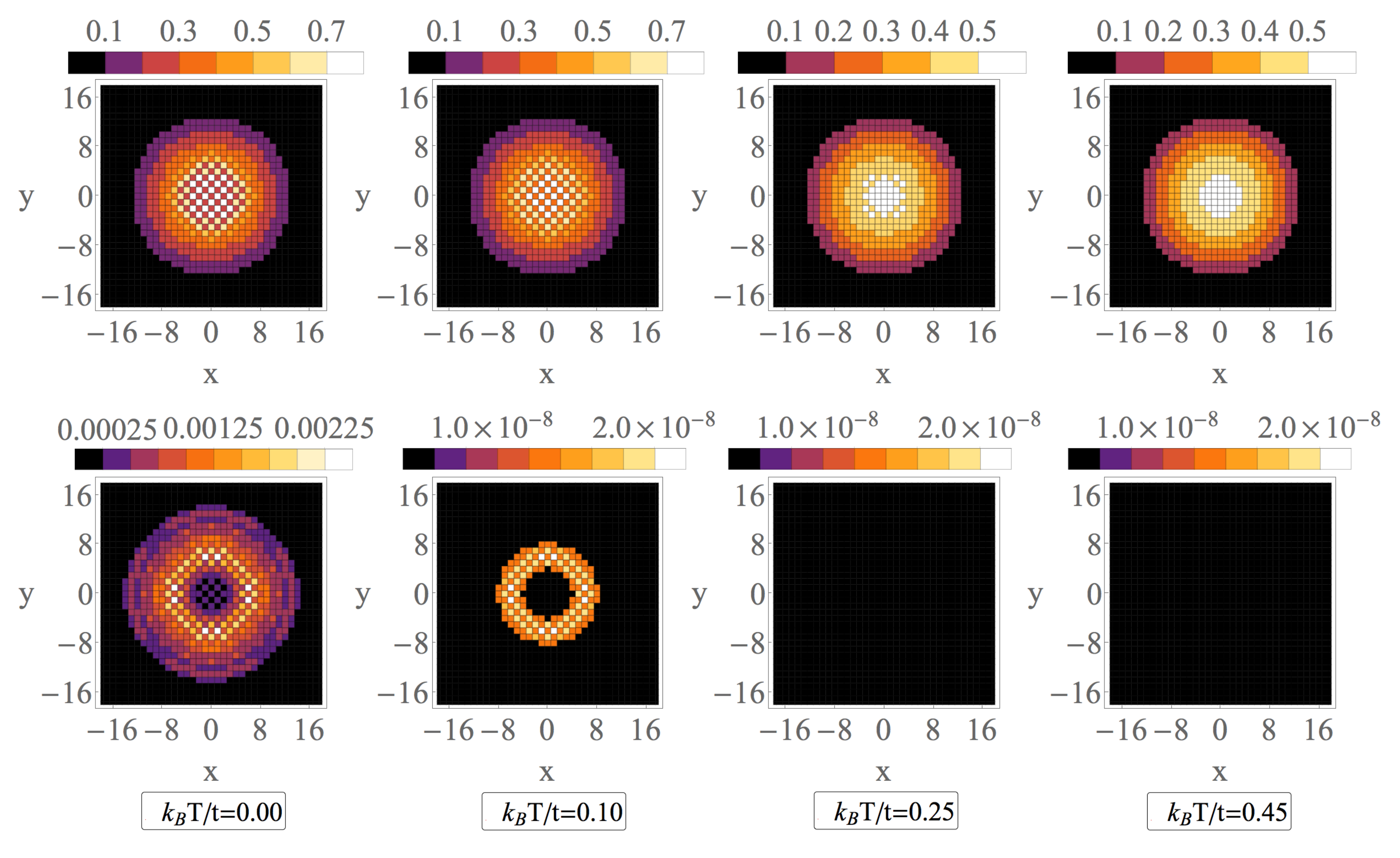} 
\end{center}
\caption{(Color online) Density order profile (top) and superfluid order parameter (bottom) as a function of the temperature. From left to right $k_B T/t=0.0, 0.10, 0.25$ and $0.45$. For $\Lambda=1.0$ and $\chi=0.3$.}
\label{Fig7}
\end{figure}

Finally, when the interlayer spacing is large enough, the intralayer repulsive interaction dominates over the interlayer attraction and the superfluid order parameter almost vanishes. This behavior is found for $\Lambda=1.0$, where pairing is inhibited. For larger values of $\Lambda$ no pairs can be formed but a DW checkerboard pattern still persists at the center of the trap (see Fig. \ref{Fig7}). This value of $\Lambda$ signals the limit from which each layer can be studied separately. The single layer model has been studied  previously considering arbitrary dipole moment orientations\cite{Gadsbolle1,Gadsbolle2}. We found that our predictions are in good agreement with those results, that is, at a given critical temperature that depends on the interaction strength, checkerboard phases for perpendicular orientation of the dipole moment emerge.  

\begin{figure}[htbp]
\begin{center}
\includegraphics[width=5.3in]{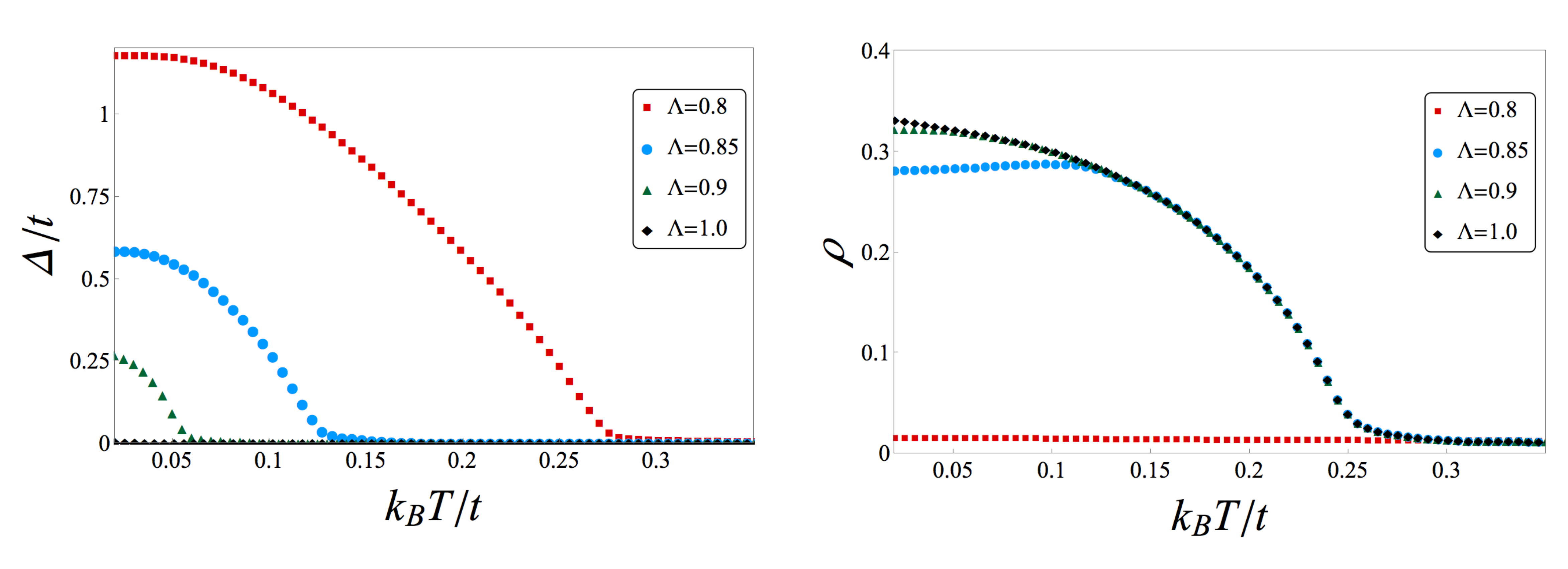} 
\end{center}
\caption{(Color online) Order Parameters obtained from BdG diagonalization. A second order continuous phase transition is shown.}
\label{Fig8}
\end{figure}

In Fig. \ref{Fig8} we plot the two global order parameters $\Delta$ and $\rho$ as a function of the temperature for values of $\Lambda$ in the region of coexistence. As can be appreciated from this figure, Bogoliubov-de Gennes diagonalization predicts continuous phase transitions for the considered model. One can observe that, for a given value of $\Lambda$, the critical temperature at which the superfluid phase emerges coincides with that at which the derivative of the DW order parameter shows a discontinuity. Numerical calculations performed for lattices of larger size ($\Omega= 2\times 57\times 57$) lead us to observe how the discontinuity of the derivative in $\rho$ and $\Delta$ at the critical temperature becomes more evident as a function of $\Omega$. Namely, the global order parameters $\Delta$ and $\rho$ change more abruptly at the critical temperature as the lattice size is increased. It is also important to stress that the referred discontinuity in the derivative becomes less evident when the interlayer spacing is increased.

\begin{figure}[htbp]
\begin{center}
\includegraphics[width=4.3in]{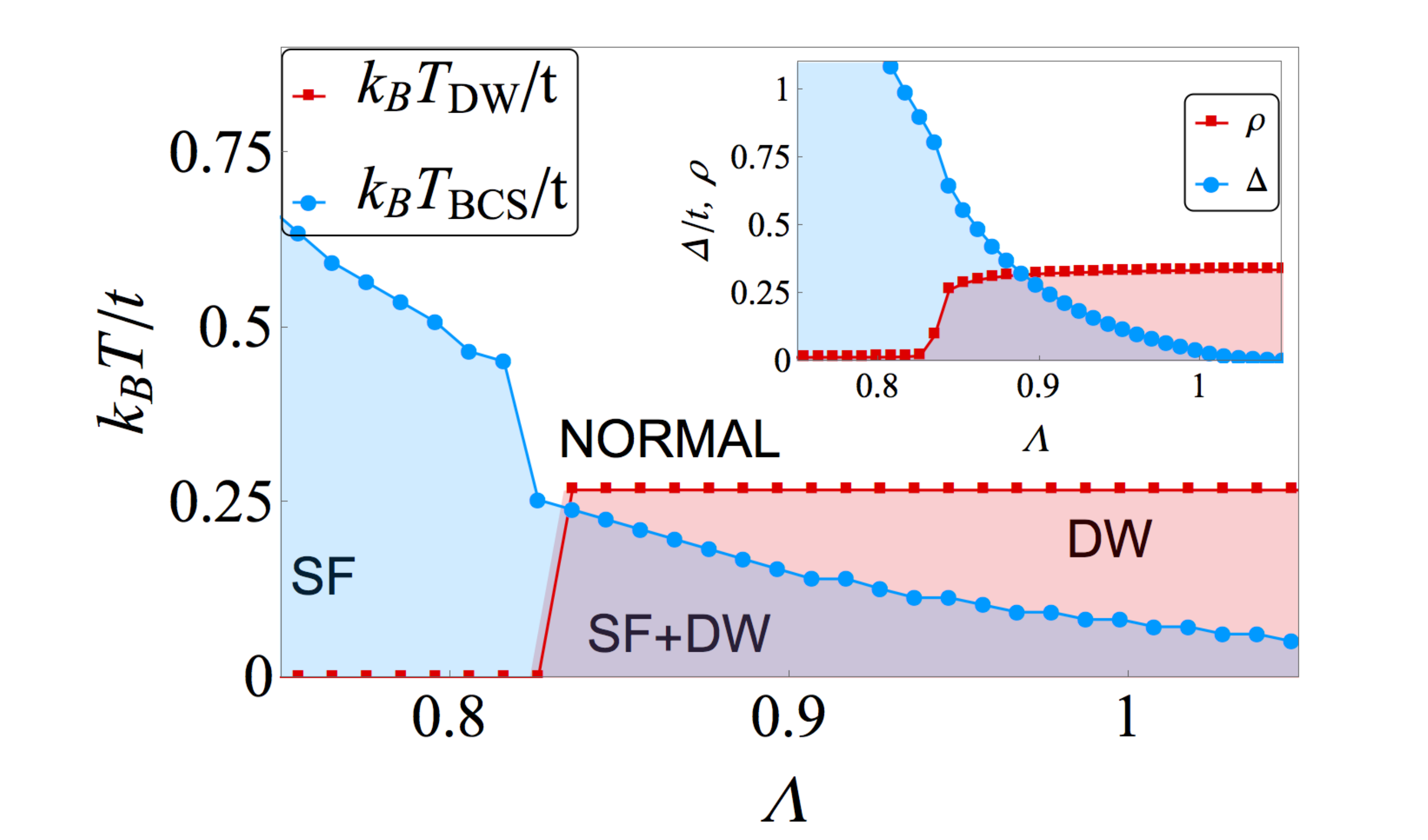} 
\end{center}
\caption{(Color online) Phase diagram for lattices of size $2\times 37 \times 37$, as a function of the dimensionless inter-layer separation $\Lambda = \lambda/a$. The interaction strength is $\chi=0.3$.}
\label{Fig9}
\end{figure}

In Fig. \ref{Fig9} we present the phase diagram of this model, obtained from Bogoliubov-de Gennes equations for finite temperatures. In the inset we show the phase diagram at zero temperature. The region of coexistence between superfluid and DW phases in both diagrams is the supersolid phase of our system. As expected, in the attractive interaction regime the superfluid phase destroys any density order pattern. When the interlayer spacing $\lambda$ becomes comparable with the lattice constant $a$, the superfluid phase and DW start to compete. For $\Lambda<0.83$ there is no formation of density order pattern, while the critical temperature of the BCS superfluid phase decrease monotonously. Close to $\Lambda\approx 0.83$ a density order pattern is formed, then the critical temperature of this phase jumps quickly to a constant value. For values of $\Lambda$ larger than $0.83$ the critical temperature of the DW phase becomes almost independent of the interlayer spacing. On the other hand, the superfluid parameter $\Delta$ suddenly decrease at $\Lambda\approx 0.83$ and then again starts to decrease monotonously. In contrast with the predictions  obtained for the single layer system where the critical temperatures may be calculated using the value of the parameters at the center of the trap, this may not be completely true for the system here studied due the possible formation of disks and rings. 

The maximum value of the critical temperature for the supersolid phase predicted by our model, considering the parameters of the current experimental systems, is $k_{B}T/t\approx 0.23$. Although such temperature is one order of magnitude smaller than those measured recently in experiments with fermonic KRb \cite{Ni} and NaK \cite{Woo}, current efforts in controlling and lowering the temperature of molecules are promising to reach such critical temperatures in the near future.

\section{Conclusion}
\label{Conclusion}
We have studied the thermodynamic phases that exhibited dipolar Fermi molecules placed at the sites of a bilayer array of square optical lattices in 2D in the presence of a harmonic confinement. Due to the nature of the dipolar interaction, where attractive and repulsive interactions are present, several phases are shown to form. While at- tractive interaction between molecules in different layers leads to predict superfluid phases, density order phases like checkerboard patterns result from the repulsive interaction. The competition between these phases gives rise to the formation of supersolid phases where both SF and DW phases coexist and spatially overlap. An exhaustive exploration of the space of parameters are summarized in the phase diagrams at zero and finite temperatures, see Fig. \ref{Fig9}. Our predictions allowed us to identify clearly the influence of the harmonic potential in the occurrence of the transitions with respect to thermodynamics in the homogeneous case reported in previous literature. The system here studied in combination with the capability of trapping Fermi molecules in optical lattices as well as the recently reported production of rovibrational and hyperfine ground state of $^{23}$Na$^{40}$K molecules constitute a promising candidate to study the competition between  BEC and BCS superfluid phases in coexistence with an ordered structure, and thus, offering the opportunity to quantum simulate a supersolid phase in ultracold experiments.

\section{Acknowledgments}

\noindent
This work was partially funded by grants IN107014 DGAPA (UNAM) and LN-232652 (CONACYT). A.C.G. acknowledges a scholarship from CONACYT.


\begin{thebibliography}{}
\bibitem{Convergence} Self-Consistent Bogoliubov-de Gennes diagonalization was iterated until convergence criteria was reached. The convergence criteria was reached when $max((|\Delta^{(m+1)}_{ij})-(\Delta^{(m)}_{ij}))|/|\Delta^{(i)}_{ij}|,|n_i^{(m+1)}-n_i^{(m+1)}|/|n_i^{(i+1)}|<10^{-5}$ using the Frobenius norm for the matrices norm.
\end{thebibliography}
\end{document}